\documentclass[prb,twocolumn,showpacs,superscriptaddress,amsmath,amssymb]{revtex4-1}
\usepackage{CJK}
\usepackage{graphicx}
\usepackage{dcolumn}
\usepackage{bm}
\usepackage{hyperref}
\usepackage{times}
\begin{document}

\begin{CJK*}{UTF8}{}
\title{Quantum order by disorder in a semiclassical spin ice}
\CJKfamily{bsmi}
\author{Yang-Zhi Chou (周揚智)} 
\affiliation{Department of Physics, National Taiwan University, Taipei 10617, Taiwan}%

\author{Ying-Jer Kao (高英哲)}
\email{yjkao@phys.ntu.edu.tw}
\affiliation{Department of Physics, National Taiwan University, Taipei 10617, Taiwan}%
\affiliation{Center of Theoretical Sciences, and Center for Quantum Science and Engineering, \\
National Taiwan University, Taipei 10617, Taiwan}%

\date{\today}

\begin{abstract}
We study the effects of quantum fluctuations in spin ice by considering an $S>1$ quantum Heisenberg model with a nearest-neighbor ferromagnetic interaction $J$ and a large non-collinear $\langle 111 \rangle$ easy-axis anisotropy $D$ on a pyrochlore lattice.  For a finite $D\gg |J|$, the low-energy physics is described by a $\langle 111 \rangle$ Ising model  with additional second- and third-neighbor exchange couplings  of order $\mathcal{O}(J^2/D)$, generated by the quantum fluctuations arising from the transverse components of the exchange coupling.  The extensive degeneracy of ground states in the $D\rightarrow \infty$ limit is lifted, and a $q=0$ ordered state is selected via the quantum order by disorder mechanism, through a first-order phase transition at low temperatures. We propose  that quantum dynamics  in spin ice can be  tuned by engineering the local single-ion anisotropy.
\end{abstract}
\maketitle
\end{CJK*}
\textit{Introduction.} ---
Spin ice materials are magnets with \textit{ferromagnetic} interactions on the geometrical frustrated pyrochlore lattice of corner-sharing tetrahedra with strong single-ion anisotropy\cite{spinice,RMPpyro, Rosenkranz00}. The presence of local, non-collinear easy axes in these materials leads to pseudo Ising spins with effective \textit{antiferromagnetic} interactions\cite{Harris}, and the system is highly frustrated. At low temperatures, the system enters  a cooperative paramagnetic state, and the magnetic moments(spins) obey the ``ice rules'', with two spins pointing into the center of each tetrahedron and two spins out. This local organizing rule gives rise to an exponentially large number of degenerate ground states, resulting in a nonzero residual entropy, and the system remains magnetically disordered.  
When  a magnetic field is applied along the [100] direction,  the spin ice enters a $q=0$ ordered state with saturated magnetization  through a topological 3D Kasteleyn transition at low temperatures\cite{100KT,Morris09}, and  the low-energy magnetic excitation involves a collection of spins on a string spanning  the entire system \cite{Jaubert09, Mono}. 

An exciting direction of research in frustrated magnetism is to understand how the extensive classical ground state degeneracy is lifted by quantum fluctuations and what types of exotic phases may emerge.
The local $\langle 111 \rangle$ anisotropy and the structure of the pyrochlore lattice in spin ice, however, precludes  the introduction of  quantum dynamics by a global magnetic field transverse to the Ising spins, although the  off-diagonal terms in the
dipolar Hamiltonian may suffice to introduce quantum dynamics in the rare-earth based spin ice materials. This issue has been  addressed  by either adding multiple-spin interactions\cite{quantumice}, or taking into account the crystal field excitations\cite{Hamid}. Here we propose a new route to explore the semiclassical  spin ice by considering the  quantum dynamics generated at finite single-ion anisotropy. 

Classical spin ice models   based on Ising-like spins  along the local $\langle 111\rangle $ easy axis\cite{DipolarIce,Siddharthan99, neuHTO, spinice} have been very successful in explaining the properties of the rare-earth based spin ice materials, such as Ho$_2$Ti$_2$O$_7$ and Dy$_2$Ti$_2$O$_7$. In these materials, the anisotropy gap $\Delta \approx 200 \sim 300$K \cite{Rosenkranz00}, is much larger than the nearest-neighbor exchange $|J| \approx $ 1K \cite{spinice,RMPpyro}, and the Ising-like models are well justified. However, one might expect in  yet-to-be-discovered spin ice materials based on transition metal ions, where the exchange is larger  and spin values are smaller, transverse quantum fluctuations due to finite anisotropy become important. Intriguing new ordered states may emerge due to the lifting of the macroscopic degeneracy via the "order by disorder" mechanism\cite{cop}.  In the spin ice, the non-collinearity of the local $\langle 111 \rangle$ easy axes generates non-trivial quantum corrections at the order of $\mathcal{O}( J^2/D)$, in the form of  second- and third-neighbor exchange couplings.  This should be contrasted with the $S>3/2$  triangular and kagome lattice antiferromagnets  with a global single-ion anisotropy, where the leading non-trivial quantum corrections are multiple-spin interactions of a higher order $\mathcal{O}(J^3/D^2)$\cite{KSL,KTSL, DPTAP}.  

\begin{figure}[b]
\centering
\includegraphics[width=0.42\textwidth,clip]{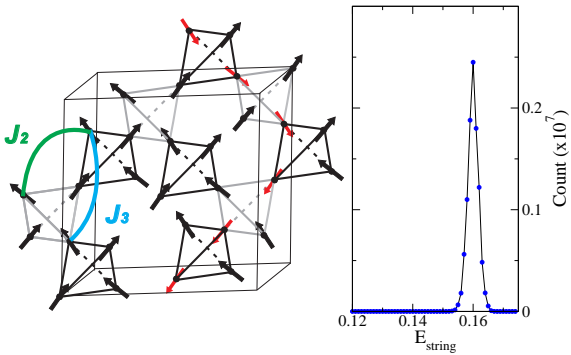}
\caption{ (Color online) Left:  Pyrochlore lattice showing $q=0$  magnetically ordered  spin structure (black narrows) and a string defect (red arrows). Second-  and third-neighbor pairs are indicated. Right: The energy distribution of the spanning string excitations from the simulation of $L=64$, starting from a $q=0$ ordered state and $10^7$ strings are generated by random walks.  }
\label{fig:100}
\end{figure}

In this paper, we consider an $S>1$ ferromagnetic Heisenberg model with a finite anisotropy $D\gg |J|$ along the local $\langle 111 \rangle$ easy axis. The low-energy physics  is described by a $\langle 111 \rangle$ Ising model with additional second- and third-neighbor exchange couplings.   We find at low temperature, the transverse quantum fluctuations due to finite anisotropy act as an order by disorder mechanism, and a $q=0$ ordered state is selected (Fig.~\ref{fig:100}).   This  state is reminiscent of the $q=0$ ordered state in Ref.~\cite{100KT}, where a 3D Kasteleyn transition is proposed. However, the specific heat in our model diverges as a power law, instead of logarithmically  when the temperature approaches  $T_c$ from above, and  the topological spanning string excitations are no longer degenerate.  We argue  that this opens up new opportunities to tune the quantum effects in spin ice by engineering the local single-ion anisotropy, and the resulting semiclassical spin ice will provide a new playground to study the quantum order by disorder phenomena.

\textit{Effective Hamiltonian.}--- We start with the quantum spin Hamiltonian ($S\ge 1$) for the nearest-neighbor ferromagnetic Heisenberg model $(J<0)$with a  local $\langle 111 \rangle$ single-ion anisotropy, 
\begin{equation}
\mathcal{H}=J\sum_{\langle ia,jb\rangle}\mathbf{S}_{ia}\cdot\mathbf{S}_{jb}-D\sum_i\left(\mathbf{S}_{ia}\cdot \hat{\mathbf{e}}^z_a\right)^2\label{Original}
\end{equation}
where $i,j$ are the FCC lattice points, $a,b$ label the sublattices inside a single tetrahedron, and the summation is over all the nearest-neighbor pairs. $\hat{\mathbf{e}}^z_a$ corresponds to the local $\langle 111\rangle$ easy axis for sublattice $a$,  and $D>0$ confines the spins to the local easy axis. For $D=0$, the model becomes a ferromagnetic Heisenberg model with a trivial ferromagnetic ground state. In the limit of $D\rightarrow \infty$, the Heisenberg model maps into a $\langle 111 \rangle$ nearest-neighbor Ising model. 
The ground states obey the 2-in-2-out ice rules for each tetrahedron with an extensive degeneracy. In the case of finite anisotropy, we expect  corrections to the classical $\langle 111\rangle $ nearest-neighbor Ising model.  In a \textit{classical} ferromagnetic Heisenberg model with finite anisotropy, a magnetically ordered ground state with a four-sublattice structure is found\cite{Champion02}. We adopt the local basis for each sublattice, and the local easy axis is chosen as the local quantization axis for the spin operators\cite{Local111}.  The Hamiltonian can be rewritten as 
 $\mathcal{H}=\mathcal{H}_0+\mathcal{H}_I$, where
\begin{eqnarray}
\nonumber\mathcal{H}_0&=&-D\sum_{ia}\left(S_{ia}^z\right)^2,\\
\nonumber\mathcal{H}_I&=&-\frac{J}{3}\sum_{\langle ia,jb\rangle }S_{ia}^zS_{jb}^z+\sum_{\langle ia,jb \rangle}\Big(a_{ab}S_{ia}^zS_{jb}^+\\
&+&a_{ba}S_{ia}^+S_{jb}^z+c_{ab}S_{ia}^+S_{jb}^-+d_{ab}S_{ia}^+S_{jb}^++h.c.\Big).\label{NLH}
\end{eqnarray}
Here $a_{ab}$, $c_{ab}$, and $d_{ab}$ are the geometrical prefactors  in the local basis expansion\cite{ChouThesis}.  We construct the effective Hamiltonian in the finite $D\gg |J|$ limit following the standard degenerate perturbation theory\cite{Ecorrelation}, and 
$\mathcal{H}_{\rm eff} = PHP+PHRHP$,  with
$P  = \sum_{\alpha} \vert \Phi_{0}\rangle \langle \Phi_{0}\vert$ and
$R = \sum_{\beta\neq 0} 
{
\vert \Phi_{\beta}\rangle 
({E_0 - E_\beta})^{-1}
\langle \Phi_{\beta}\vert
}
$, 
where $E_\alpha=\langle \Phi_{\alpha}\vert \mathcal{H}_{0}\vert  \Phi_{\alpha}\rangle$. Here $\{\vert \Phi_{\alpha} \rangle\}$ are eigenstates of $\mathcal{H}_0$.
In the large $D$ limit, the ground states of $\mathcal{H}_0$ contain two maximum spin configurations $|\pm S\rangle$ at each site.

In contrast to the 2D models with a strong  \textit{global} easy-axis anisotropy, the non-collinear local $\langle 111 \rangle$ easy axes introduce extra terms previously absent in the Hamiltonian, such as  $S_i^+S_j^+$, $S_i^-S_j^-$, $S_i^+S_j^z$, $S_i^-S_j^z$, etc. 
For $S=1$, the effective Hamiltonian will include transverse terms since it is possible to bring a spin state from $|\pm1\rangle$ to $|\mp1\rangle$ via two spin raising or lowering operations. For $S>1$,  it merely consists of logitudinal terms since the net effect of the perturbation cannot alter the spin states $|\pm S\rangle$ at each site.  In the following we will restrict our discussion to $S>1$. 
In the second order term of the effective Hamiltonian, 
$S^+$ and $S^-$ should appear in pairs on each site for non-vanishing matrix elements. Thus, nontrivial second- and third-neighbor interactions result from combinations of $S_{ia}^zS_{kc}^+$ and $S_{kc}^-S_{jb}^z$ in Eq.~(\ref{Original}). Such perturbation generates an effective $S_{ia}^zS_{jb}^z$ exchange interaction with a virtual spin raising and lowering process at an intermediate site $k$.  All other second order terms merely renormalize the nearest-neighbor exchange. Therefore,  extra exchange interaction  in the effective Hamiltonian are couplings between sites linked by two intermediate nearest-neighbor bonds (Fig.~\ref{fig:100}).

\textit{Ising equivalents.}--- We can map the Heisenberg operators into an effective Ising Hamiltonian. There are four operator equivalents\cite{ChouThesis}: (1) $S^z_{ia}=S\sigma_{ia}^z$, (2) $S_{ia}^\pm S_{ia}^\mp= 2S\times(1\pm\sigma_{ia}^z)/2$,  
(3) $S_{ia}^{-}S_{jb}^{+}S_{ia}^{+}S_{jb}^{-}+S_{ia}^{+}S_{jb}^{-}S_{ia}^{-}S_{jb}^{+}= 4S^2\times(1-\sigma_{ia}^z\sigma_{jb}^z)/2$ 
(4) $S_{ia}^{-}S_{jb}^{-}S_{ia}^{+}S_{jb}^{+}+S_{ia}^{+}S_{jb}^{+}S_{ia}^{-}S_{jb}^{-}= 4S^2\times(1+\sigma_{ia}^z\sigma_{jb}^z)/2$, where $\sigma^z=\pm 1$. Using these operator equivalents, the new effective $\langle 111 \rangle$ Ising model is written as
\begin{eqnarray}
\nonumber\mathcal{H}_{\rm eff}&=&\mbox{const.}+J_1\sum_{\langle ia,jb\rangle}\sigma_{ia}^z\sigma_{jb}^z\\
&+&J_2\sum_{\langle \langle ia,jb\rangle\rangle}\sigma_{ia}^z\sigma_{jb}^z+J_3\sum_{((ia,jb))}\sigma_{ia}^z\sigma_{jb}^z,\label{HeffL}
\end{eqnarray}
where $\langle\langle i,j\rangle\rangle$ and $((i,j))$ are the second and third neighbor pairs.
The $J_1$ and $J_2$ both are antiferromagnetic, with  
 $J_1=-\frac{JS^2}{3}+\frac{J^2}{D(2S-1)}\left(\frac{4S^3}{9}-\frac{S^2}{12}\right)>0$ for $S>1$ and $J_2=\frac{J^2}{D(2S-1)}\frac{2S^3}{9}>0$.  $J_3$ is ferromagnetic, with $J_3=-\frac{J^2}{D(2S-1)}\frac{4S^3}{9}=-2J_2<0$    \cite{Chern08note}. 

 For a given site, there are six $J_1$ bonds, twelve $J_2$ bonds, and six $J_3$ bonds with an intermediate site  (Fig.~\ref{fig:100}). 
When $D\rightarrow \infty$, the model reduces to the $\langle 111 \rangle$ nearest-neighbor spin ice model. The appearance of $J_2$ and $J_3$ is due to non-trivial quantum fluctuations, and we analyze their effects on the highly degenerate spin ice states via both mean field theory and classical Monte Carlo simulations. In the following, we choose a representative value of $J_2/J_1= 0.005$, which corresponds to $|J/D|\approx 0.015$ in the large $S$ limit, and all the energies are in units of $J_1$.

\begin{figure}[t]
\centering
\includegraphics[width=0.42\textwidth,clip]{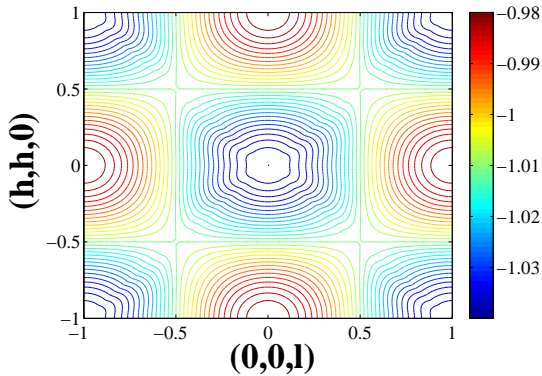}
\caption{(Color online) Lowest	eigenvalue of the interaction matrix $\mathcal{J}^{ab}(\mathbf{q})$ at wave vectors $\textbf{q} = 2\pi(h, h , l)$. The minimum occurs at $\mathbf{q}=(0,0,0)$.}
\label{fig:Eigenvalue}
\end{figure}

\begin{figure}[tb]
\centering
\includegraphics[width=0.42\textwidth, clip]{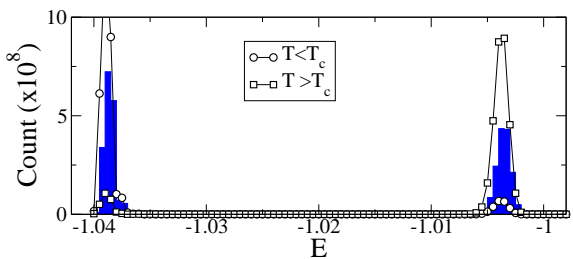}
\caption{The energy histogram for $L=8$ near the  transition temperature. $J_2=0.005J_1$ and $J_3=-2J_2$. The double peak structure in the histogram indicates a first order phase transition.}
\label{fig:Hist}
\end{figure}
\textit{Mean-field theory.}--- To study possible ordering, we first analyze the effective Hamiltonian Eq.~(\ref{HeffL})  by the  mean-field theory\cite{pyrochloreMFT,NeuMFT}. The free energy  up to the quadratic order at temperature $T$ is given by
$\mathcal{F}=\frac{1}{2}\sum_q\sum_{ab}\left(-\mathcal{J}^{ab}(q)+T\delta_{a,b}\right)m^{a}_q m^{b}_{-q}+O(m^4),$
where $m^{a}_q=\frac{1}{N}\sum_i e^{i\mathbf{q}\cdot\mathbf{r}^{a}_i}\sigma_{ia}$, and $a,b$ are sublattice indices.  The matrix $\mathcal{J}^{ab}(\mathbf{q})$ is the Fourier transform of the interaction matrix $\mathcal{J}^{ab}_{ij}=\mathcal{J}(\mathbf{r}_i^a-\mathbf{r}_j^b)$. The lowest eigenvalue of $\mathcal{J}$ is associated with the first ordered mode of the model.  In the infinite anisotropy limit, the lowest eigenvalues form a $\mathbf{q}$-independent flat band, which corresponds to the macroscopically degenerate ground states and  no preferred ordering is selected. This degeneracy is lifted in the presence of $J_2$ and $J_3$, and the lowest eigenvalue minimum is located at $\mathbf{q}=(0,0,0)$(Fig.~\ref{fig:Eigenvalue}), and the system develops a $q=0$ long-range order. 


\begin{figure}[tb]
\centering
\includegraphics[width=0.42\textwidth, clip]{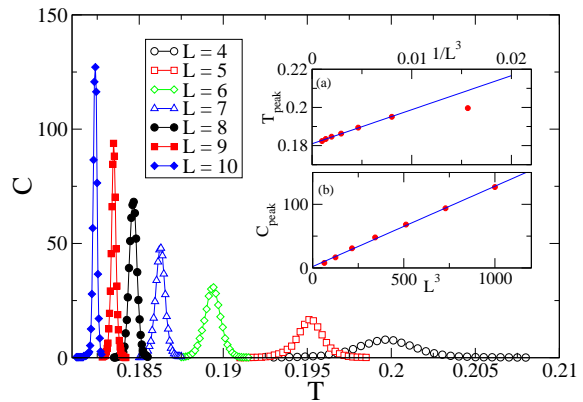}
\caption{(Color online) Variation of specific heat as a function of temperature for several different sizes. Insets: (a)  Specific peak versus $L^3$. A clear extensive behavior is shown in the linear fit. (b)  The peak temperature versus $1/L^3$. The transition temperature in thermodynamic limit is extrapolated to be $T_c^\infty=0.181$. The behavior is consistent with the properties of a first-order phase transition. }
\label{fig:CT}
\end{figure}

\begin{figure}[th]
\begin{center}
\includegraphics[width=0.42\textwidth,clip]{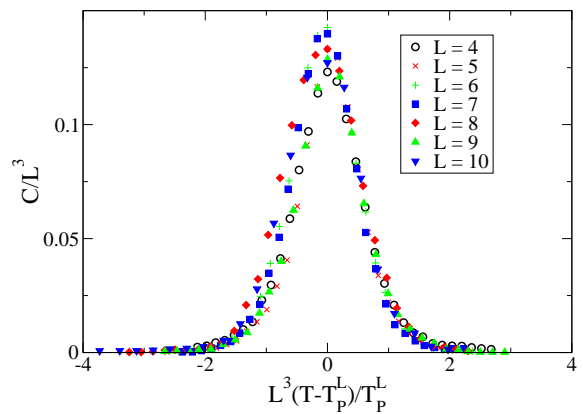}
\caption{ (Color online) Scaling of the specific-heat data. $T^L_p$ corresponds to the temperature of the specific heat maximum for a given size $L$. \vspace{-0.2in}}
\label{fig:scaling}
\end{center}
\end{figure}

\textit{ Monte Carlo Simulation.}--- With the information of a possible $q=0$ order from the mean-field analysis, we study the effective Hamiltonian Eq.~(\ref{HeffL}) using Monte Carlo simulations. The simulation is done on the pyrochlore lattice with periodic boundary conditions measuring $L$ cubic unit cells in each direction, which amounts to a total of  $N=16\times L^3$ spins. We perform our simulations using the parallel tempering algorithm\cite{MCPT},  so that the simulation can reach ergodicity  more efficiently at low temperatures. The simulations are conducted in parallel at a series of temperatures, and  swaps of configurations between these parallel simulations are proposed, allowing the low temperature system of interest to escape from local free energy minima where it might otherwise be trapped. In addition,  the configurations still favors ice rule for $J_1\gg J_2$ in each serial run, so  we also employ loop updates to avoid the ice-rule breaking energy barrier\cite{MelkoReview}.
In our simulations, for $L=4$ to 7, we carry out 2000 configuration swaps and 200 loop updates between swaps  in both equilibrium and sampling processes;  for $L=8$ to 10, 4000 configuration swaps and 200 loop updates between swaps during equilibration and 2000 swaps and 200 loop updates between swaps in sampling are carried out.  Figure~\ref{fig:Hist} shows the energy histogram for $L=8$ at temperatures near the transition. The double peak feature at the transition  indicates the coexistence of two distinct phases at $T_c$.  Figure~\ref{fig:CT} shows the temperature variation of the specific heat for various lattice sizes. As the size increases, the peaks grow and peak widths decrease. There is also a shift in the temperature corresponding to the specific heat peak $T_p$. The specific heat in a first order transition should be extensive, and diverges  in the thermodynamic limit. We fit the size dependence of the specific heat by $C=2.14 L^3+0.126$ (inset (a) of Fig.~\ref{fig:CT}). On the other hand,  the peak temperature $T_p^L$ for a  given size $L$ can be fit to $T^L_p=T_c^\infty+1.78/L^3$ with $T_c^\infty=0.181$ (inset (b) of  Fig.~\ref{fig:CT}). The above two relations suggest that it is possible to perform scaling analysis on the specific heat data using $C$ and $\Delta T=T-T_L^p$. Figure~\ref{fig:scaling} shows the  scaling of the specific heat data, and  the scaling works reasonably well. We conclude that this transition to the $q=0$ ordered state is a first order phase transition\cite{FSS}.

At low temperatures, the ground state corresponds to a $q=0$ ordered state and carries saturated magnetization toward one of the $[100]$ directions(Fig.~\ref{fig:100}). The ground state energy can be computed exactly, $E_g/N=-J_1-2J_2+3J_3=-J_1-8J_2$. In the ice-rule states, the low energy excitation is  a collection of  flipped spins on a string, as a single spin-flip  costs higher exchange energy of order $\mathcal{O}(J_1)$.  The spin ice state and the $q=0$ state both obey the local 2-in-2-out constraint but their excitations  show distinct topology. In the spin ice state, the strings form loops of finite lengths, while in the $q=0$ ordered state, the string will span the entire system.  These spanning strings excitations carry  energy and entropy proportional to the segment length of the string.  At high enough temperature, such excitations will lower the free energy and the ordered state is destroyed. This is similar to the case of the spin ice in a $[100]$ field\cite{100KT}, where a 3D Kasteleyn transition into a $q=0$ ordered state at low temperature is proposed.
In our model, however, the specific heat diverges as a power law when the temperature approaches $T_c$ from above; while in the case of a spin ice in [100] field, it diverges logarithmically. We note that in our model the exact degeneracy of the spanning string excitations in Ref.~\cite{100KT} is lifted and the  excitation energy depends on the path the string traverses. Figure~\ref{fig:100} shows the segment energy distribution of the spanning string excitations from the simulation of $L = 64$, starting from a $q=0$ ordered state and a total of $10^7$ strings are generated by random walks\cite{100KT,Morris09}. A broad distribution is clearly observed due to the path dependence of the energy in the string creation.

\textit{Conclusion.}--- We propose that  a semiclassical spin ice with finite anisotropy provides a new playground to study the quantum order by disorder phenomena. Transverse quantum fluctuations, which can be tuned by engineering the local single-ion anisotropy, generate additional second- and third-neighbor exchange interactions of order $\mathcal{O}(J^2/D)$ in the $\langle 111 \rangle $ Ising model. These new interactions lift the extensive ground state degeneracy in the infinite anisotropy limit and select a six-fold $q=0$ ordered ground state carrying saturated magnetization toward one of the $[100]$ directions. Although the topological characteristics of the string excitations are similar to those in Ref.~\cite{100KT}, the critical behavior is quite different due to the path dependence of the spanning string energy.  Interesting questions remain on the effects of an external magnetic field and dilution\cite{dilute}, which requires further study.

\begin{acknowledgments}
We are grateful to R. Melko, M. J. P. Gingras and P. Fulde for useful discussions.  
We thank the NCHC of Taiwan for the support of high-performance computing facilities.This work was supported by the NCTS and the NSC of Taiwan through Grant Nos. NSC-97-2628-M-002-011-MY3, NSC-98-2120-M-002-010-, and by NTU Grant Nos. 97R0066-65 and 97R0066-68.
\end{acknowledgments}

\bibliography{MyBib}

\begin{thebibliography}{10}
\bibitem{Local111}%
  \BibitemOpen
  \bibinfo {note} {The choices of the local $z$-axes for each sublattice are:
  $\hat{\mathbf{e}}^z_1=\frac{1}{\sqrt{3}}(-1,-1,-1)$,
  $\hat{\mathbf{e}}^z_2=\frac{1}{\sqrt{3}}(1,1,-1)$,
  $\hat{\mathbf{e}}^z_3=\frac{1}{\sqrt{3}}(-1,1,-1)$, and
  $\hat{\mathbf{e}}^z_4=\frac{1}{\sqrt{3}}(1,-1,-1)$}%
  \bibAnnoteFile{NoStop}{Local111}%
\end{thebibliography}

\end{document}